\documentclass[mathleft
]{an}
\usepackage{graphicx}
\usepackage{times}
\usepackage{natbib}
\bibpunct{(}{)}{;}{a}{}{,}

\newcommand{\degree}{\ensuremath{^\circ}}

\newcommand{\aap}{    {Astron. Astrophys.}}

\newcommand{\nat}{    { Nature}}
\newcommand{\solphys}{{Solar Phys.}}
\newcommand{\ssr}{    { Space Sci. Rev.}}

\begin{document}

\Yearpublication{2012}%
\Yearsubmission{2012}%
 \DOI{This.is/not.aDOI}%

\title{The Current State-Of-The-Art In Active Region Seismology}

\author{Hamed Moradi\thanks{Corresponding author: \email{hamed.moradi@monash.edu}}
}
\titlerunning{The Current State-Of-The-Art In Active Region Seismology}
\authorrunning{Hamed Moradi}
\institute{Monash Centre for Astrophysics, School of Mathematical Sciences, Monash University, Victoria 3800, Australia}

\received{15 Aug 2012}
\accepted{17 Oct 2012}


\abstract{%
Helioseismology is the study of the variations in the internal structure and properties of the dynamics of the Sun from measurements of its surface oscillations. With the 2010 launch of the Solar Dynamics Observatory (SDO) we are undoubtedly approaching a new dawn for local helioseismology, as the extent and quality of raw surface oscillation data has never been better. However, advances in theory and modelling are still required to fully utilise these data, especially in magnetic active regions and sunspots, where the physics is poorly understood.}

\maketitle

\section{Introduction}

Understanding the origin of the Sun's magnetic field is arguably the most important topic in solar/stellar physics today. Sunspots and starspots are the most visible manifestations of solar/stellar magnetic activity, with sunspots tending to appear at well-defined latitudes on the Sun, varying with the 11-year solar cycle. Our comprehension of, and capability to predict, the 11-year activity cycle of the Sun is far from being complete however; as suggested by the recent unusually long solar minimum of Cycle 23. Understanding the mechanism of the solar dynamo is therefore of the utmost importance, not only to be able to explain this collective behaviour in the Sun, but to also explain the fundamental physical processes of the space environment from the Sun to Earth, to other planets, and beyond to the interstellar medium. 

Sunspots can provide us with important clues on how the solar dynamo operates, therefore a comprehensive discernment of their subsurface evolution, structure and dynamics is essential in order to establish accurate physical relationships between internal solar properties and magnetic activity in the photosphere. Helioseismology is the only tool available that can probe the structure and dynamics of the Sun through the study of acoustic oscillations.

\subsection{Local Helioseismology}

Helioseismology is a diagnostic tool that allows us to probe the solar parameters to compare with theory and observation. Over the last fifteen years, considerable progress has been made in the understanding of the solar interior, mainly as a result of the precise determination of the frequencies of millions of individual global modes of oscillation using observations from the Michelson Doppler Imager (MDI; \citealp{scherreretal1995}), on board the Solar and Heliospheric Observatory (SoHO) spacecraft, and ground-based networks of telescopes. Measurements of flows and temperature variations in the solar interior have already led to an increased understanding of solar activity; for example, the detection of subsurface meridional flow (e.g., \citealp{gilesetal1997}) has had a major impact on theories of the solar dynamo. 

Local helioseismology has also shown that wave propagation is different in sunspots than in the quiet Sun \citep{gizonetal2009,moradietal2010}, with the wave travel-time shifts in the vicinity of sunspots typically interpreted as arising predominantly from magnetic fields, flows, and local changes in sound speed. The magnitude of the travel time shifts also appear to acutely depend on the details of the data analysis, the wave travel distance, and frequency \citep{cr2007,bb2008,tz2008,mhc2009}. 

Various local helioseismic techniques exist to analyse the oscillations. For example, Fourier-Hankel analysis has been used successfully in the past for studying the interaction of acoustic waves with sunspots \citep{bdl1987}. This process decomposes the solar oscillation signal, observed in an annulus around a sunspot, into inward and outward propagating wave modes. Another technique is the ring-diagram method \citep{hill1988}. Ring-diagrams are local spatial power spectra at constant frequency. Cuts at constant frequency through the 3D power-spectra reveal nested ellipses or rings that change shape and shift centre under the influence of alterations to the solar medium. Time-distance analysis is another method which computes the travel time of a wave packet travelling between two points on the solar surface \citep{duvalletal1997}. Finally, there is the method of helioseismic holography (acoustic imaging), which reconstructs the subsurface wave fields by propagating waves either forward or backward in time \citep{lb1997,changetal1997}.

In this review, I present a brief overview of the current state-of-the-art in the local helioseismology of magnetic activity, discussing a number of recent findings and some recent advances in theory that will allow for new insights.

\section{Local Helioseismology of Magnetic Activity}

\subsection{Active Region Emergence}

Since its inception, one of the core scientific goals of helioseismology has been to accurately detect the emergence of magnetic activity below the surface. The realization of this goal would obviously have considerable implications for improving the accuracy of space weather forecasts and predictions. 
There are a number of recent results from local helioseismology that are beginning to show promising results on this matter. 

In a series of studies, \cite{kommetal2007,kommetal2009,khh2009} and \cite{jainetal2012} have used ring diagrams to analyse the subsurface flow systems of a large sample of active and quite-Sun regions. Their studies have revealed that flux emergence is typically accompanied by a number of characteristics, namely: up-flows, an increase in subsurface zonal flow velocity, and an increase in vorticity of subsurface flows, while essentially the opposite is observed for decaying flux. 

In another study, \cite{izk2011} have used ``deep-focus'' time-distance helioseismology to detect strong travel-time reductions ($\sim$ 12 -- 16 seconds) at great depths ($\sim$ 42 -- 75 Mm), some 1 -- 2 days before high peaks in the photospheric magnetic flux rate, for a number of different active regions. These results are somewhat controversial however, as the amplitude of these reported travel-time perturbations are significantly larger than theoretical estimates based on numerical simulations of emerging flux tubes \citep{bbc2010}. An independent analysis of the same active regions by \cite{braun2012}, using helioseismic holography, failed to confirm the reported travel-time anomalies of \cite{izk2011}, reporting time shifts of order 1 -- 2 s for the same set of active regions. 

While it has been suggested that subtle differences in the measurement geometries/averaging schemes may have played a role in the diverging results \citep{izk2012}, the root cause(s) behind why such a large discrepancy exists between the two helioseismic methods, and between the time-distance results and numerical simulations of emerging flux, is currently an open question. \cite{braun2012} makes a salient point in that the issue is perhaps best resolved through blind tests using artificial data. 
  
\subsection{Sunspot Seismology}

There have been many (at times conflicting) conclusions drawn about sunspot and active region structure from local helioseismology (see \citealp{gbs2010}). The predominant approach has been to directly interpret helioseismic measurements and infer the interior properties based on models of how perturbations affect these metrics. The physics of propagating waves and their interactions with magnetic fields are however non-trivial, with mode conversion, which is the exchange of energy between the different modes of oscillation that are supported by a magnetized plasma (fast and slow magnetoacoustic-gravity waves, and Alfv\'en waves) appearing to play a key role in helioseismic measurements. Mode conversion is typically strongest in the region where the Alfv\'en wave speed ($c_a$) coincides with the sound speed ($c_s$) (e.g. \citealp{cally2005,crouchetal2005,cally2006,cally2007,sc2006}). 

The strong magnetic fields which permeate the solar surface also produce acute surface phase perturbations that obscure helioseismic observations, particularly in sunspot penumbrae \citep{lb2003,lb2005,schunkeretal2005}, thereby making magnetic effects much more complicated and harder to disentangle from other non-linear effects such, as radiative transfer effects on Doppler measurements \citep{rajaguruetal2007, rajaguru2012}. 

Interpreting helioseismic results are often further complicated when considering the effects of filtering on the data, a particularly poignant issue in time-distance helioseismology, where typically either phase-speed filtering or ridge-filtering is employed. Phase-speed filtering was originally introduced by \cite{duvalletal1997} in order to improve time-distance measurements at small distances. This type of filtering roughly isolates modes that travel a similar path in the solar interior. In ridge filtering, modes with a common radial order (evident as the ``ridges'' in a power spectrum) are selected \citep{dg2000}. 

In an important study, \cite{bb2008} showed that the type of filtering employed plays a significant role in both the sign and magnitude of measured travel time perturbations in and around active regions, as phase-speed filters can produce anomalous sign-changes as the perturbed ridge power falls inside or outside of the filter. This in turn has significant implications for the inferred magnitude and lateral extent of subsurface structure and flows inversions derived from linear inversions of phase-speed filtered helioseismic data (see e.g., \citealp{gizonetal2009,moradietal2010}).    

\subsubsection{Sunspot Flows}

An example of how sensitive inversion results are to the type of data filtering employed can be seen in recent inferences regarding the nature of near-surface flows in and around sunspots. In the photosphere, sunspots are typically surrounded by diverging horizontal outflows, termed ``moat flows''. With amplitudes of several hundred m/s,  they can be seen by local correlation tracking \citep{ns1988} of the proper motions of granules \citep{sr2007, vdetal2007,vdetal2008,bm2010} and moving magnetic features \citep{sheeley1969,bl1988,gizonetal2009}, as well as through Doppler-shift measurements \citep{sheeley1972}. The moat flow also been detected by local helioseismology \citep{gdl2000}. 

In a recent result, the horizontal and vertical extent of the near-surface flows associated with the sunspot in AR 9787 was characterised by \cite{gizonetal2009}.  Using linear inversions of ridge-filtered time-distance travel times (in conjunction with Born-approximation sensitivity kernels of \cite{bg2007}), they detected a strong outflow of several hundred m/s extending beyond the penumbra of the sunspot, down to a depth of around 4.5 Mm. These results were later confirmed by \cite{moradietal2010} using high resolution ring diagram analysis. More recent 3D ring-diagram inversions by \cite{fht2011} (using the Born-approximation kernels of \cite{birchetal2007}), and inversions using helioseismic holography \citep{braunetal2011} have also confirmed similar near-surface outflows associated with sunspots. These results contrast strongly with those of \cite{zkd2001}, who used linear inversions of phase-speed filtered time-distance travel times and ray-approximation sensitivity kernels, to infer near-surface inflows (i.e., flows in the opposite direction to the observed moat flow) associated with sunspots. 

\subsubsection{Sunspot Structure}

Almost all of the various local helioseismic diagnostic tools have also been used to infer changes in the wave speed in sunspots (e.g., \citealp{fbc1995,kds2000,bab2004,cbk2006,znt2007,bbs2011}). But despite the long history of work on this topic, there is still no general agreement on the subsurface structure of sunspots. \cite{gizonetal2009} and \cite{moradietal2010} have shown that inferences made by time-distance and ring-diagram analysis with regards to subsurface wave-speed structure are surprisingly contradictory, with the inversions producing subsurface wave-speed profiles of opposite signs and different amplitudes. 

The fact that neither method fully accounts for the details of the measurement procedures (e.g., the effects of phase-speed filtering), and that both inversion methods use sensitivity functions that do not explicitly include the direct effects of the magnetic field, are perhaps the most poignant factors which contribute to such stark disagreements between the two methods. However, when \cite{moradietal2010} and \cite{gbs2010} compared the linear inversion results with a number of phenomenological and purely numerical models of subsurface wave-speed structure (e.g., \citealp{fbc1995,crouchetal2005,rsk2009,cameronetal2011}), they found that the ring-diagram inversions and the phenomenological/numerical models were actually in agreement -- to an extent --  displaying positive wave-speed perturbations down to a depth of $\sim 2$ Mm.  

As solar imaging hardware becomes increasingly sophisticated, with the Helioseismic and Magnetic Imager (HMI; \citealp{scherreretal2012}) onboard SDO providing continuous Doppler and magnetic images covering the full solar disk with fourfold better spatial resolution than its predecessor instrument (SoHO-MDI), the need for refined numerical modelling/computational techniques becomes apparent to realize the full potential that these observations offer.

\section{Numerical Modelling}

\subsection{Forward Modelling}
In the last few years, several advanced 3D (mainly linear) parallel codes have been developed to conduct computational simulations of magnetohydrodynamic (MHD) waves in sunspots, where strong magnetic fields and severe gravitational stratification produce complex behaviours in the waves we use to probe the interior (e.g., \citealp{hanasoge2008,cgd2008,khomenkoetal2009,shelyagetal2009,pk2009}). Through the process known as \emph{forward modelling}  --  constructing computational models that mimic wave propagation through sunspots and matching the resultant wave statistics with observations -- we can refine the helioseismic inferences that are currently fooled by these wave interaction and conversion processes. The major advantage of this approach is that it permits the study of wave propagation by providing the freedom to choose various background models, without the complications of solving for convection. This substantially reduces the computational expense for detailed statistical studies. Interpretations of available results from the SoHO and SDO missions can thus be re-evaluated with these studies.  

A successful example of this approach is from the analysis undertaken by \cite{cameronetal2011}, who model the propagation of $f$, $p_1$ and $p_2$ planar wave packets through a shallow, semi-empirical, non-magnetohydrostatic (MHS) sunspot model of AR 9787. In order to assess the validity of the underlying sunspot model, the authors compared the simulated wave field to the observed SoHO-MDI cross-covariance functions. They found a good quantitative match between the phase shifts and amplitudes of the simulations and actual observations of AR 9787, which is strong evidence that the wave signal is dominated by the near-surface layer of the sunspot. 

A recent (observational) study of sunspot-induced acoustic wave scattering by \cite{zcy2011} also tends to favour this assertion. Using data from the HMI instrument, they analysed the wavefunctions of scattered waves with the incident waves for the $f$ to $p_5$ mode for two sunspots.  They observed that the magnitude of the scattered wave relative to that of the incident wave, decreases from the $f$- to the $p_5$ mode, indicating that the region producing the scattered wave is shallower than the depth of the $f$-modes.

\subsection{Inverse Modelling}
The next step in the numerical modelling process is to identify a clear and efficient way to alter sunspot models and the flows around them in order to better match observations. This will entail the solving of an inverse problem, which for local helioseismology is to use measured helioseismic signatures (e.g., travel times) to estimate physical conditions in the solar interior. While recent progress has been made in improving the traditional linear sensitivity kernels and inversion algorithms (e.g., \citealp{svandaetal2011,jetal2012}), these methods have traditionally relied on background models being translationally invariant (e.g. \citealp{gb2002}), which is problematic (as discussed previously), since the nature of waves in solar active regions is significantly altered as they are converted into magneto-acoustic waves. Given the mathematical difficulties in modelling such complex phenomena, current linear inversions do not take the effects of the sunspot magnetic field into account, nor the possibility that such wide deviations from the quiet Sun introduce a non-linear relation between observed travel-time shifts and causative agents. 

A significant breakthrough has been made on this front however, with the development of an improved inverse method for helioseismology. The \emph{adjoint method}, employed by \cite{hanasogeetal2011, hanasogeetal2012}, is a form of partial-differential-equation constrained optimization that describes a means of extracting derivatives of a misfit function with respect to model parameters through finite computation. The technique allows for the computation of sensitivity kernels around arbitrary background models, thereby paving the way for non-linear iterative inversions. On another front, \cite{crouchetal2011} have managed to successfully invert for the sound-speed and magnetic field in a simple 1D model.

\subsection{Numerical Issues}

While advances in numerical modelling techniques has allowed us to make some significant inroads into a number of outstanding issues in the local helioseismology of magnetic activity, they are not without their own problems. In this section, I discuss a number of the more pressing numerical issues, affecting both forward and inverse modelling, that need to be addressed in order for us to have more confidence in the numerical simulations.   

i) Convective instability: As the Sun is highly convectively unstable in the outer 2-3\% layers, simulating small perturbations around this state is not feasible (a particularly pertinent issue for the linear wave propagation codes).  A somewhat passive approach which has been commonly adopted by the helioseismology community, is to slightly alter the background model (typically Model S; \citealp{cdetal1996}) in order to keep the Brunt-V\"{a}s\"al\"a frequency positive \citep{hanasogeetal2006,pk2007a,shelyagetal2009,schunkeretal2011}. 

While this method produces solar-like power spectrums, it is far from an ideal solution, as changing the background in such a manner effectively shifts the eigenfunctions and eigenfrequencies of the resonant modes, and may also affect the seismic reciprocity of the system \citep{dt1998}. A clear idea of how to successfully deal with the convective instability associated with the outermost layers currently has not been explored.
 
ii) Sunspot models: Numerical models of sunspots are a necessity for both forward and inverse modelling. As detailed in \cite{moradietal2010}, there is certainly no shortage of models available for helioseismologists to play with, which in itself is not a disadvantage, as examining a number of different models is essential to test the robustness of helioseismic inferences. Ideally, sunspot models should be constrained by surface observations and constructed with the subphotospheric layers in force balance and the atmospheric region force-free. 

On the other hand, fully realistic magnetoconvective simulations of sunspot structure are now a reality (e.g., \citealp{heinemannetal2007,rsk2009,rempeletal2009,rempel2012}), but currently too cost prohibitive to be employed for the exhaustive statistical studies which are inherently required for computational helioseismology. Nonetheless, such simulations are a valuable tool for improving both forward and inverse modelling techniques. A recent analysis by \cite{braunetal2012} has shown a very good quantitative agreement between the travel-time shifts of a magnetoconvective simulation and those of two actual sunspots measured using helioseismic holography.
 
iii) Alfv\'en wave speed: One of the biggest difficulties in simulating 3D MHD wave propagation in the Sun is due to the excessively large value that $c_a$ attains above the surface, approaching values of several hundred km/s. This causes the wavelengths of both the fast and Alfv\'en waves to become rather large, resulting in an extremely stiff numerical problem for the simulation codes, with the simulation timestep ($\Delta t$) being highly constrained by the Courant-Friedrichs-Lewy (CFL) condition (typically $\Delta t \sim \Delta z/c_a$; where $\Delta z$ denotes the vertical grid resolution). As such, in order to simulate artificial data sets on the time scales required for computational helioseismology (in a finite amount of time), a number of wave propagation codes apply somewhat simplistic approximations to treat the overlying model-sunspot atmosphere  (e.g., \citealp{rsk2009,hanasoge2008,cameronetal2011,braunetal2012}). The most common approach has been to apply a $c_a$ ``limiter'' to moderate the action of the Lorentz forces when $c_s$/$c_a$ becomes unreasonably big, with $c_a$ being typically (smoothly) capped at values of between 20-60 km/s  above the surface. 

The physical implications of artificially limiting $c_a$ on the seismology are not well understood however. While some choices for these limiters have been discussed before (e.g., \citealp{cgd2008,cameronetal2011,rsk2009,braunetal2012}), no serious effort has yet gone into understanding their impact on the near- and far-scattered wave-field. This is a particularly pertinent issue, given the critical role that the exponentially increasing $c_a$ appears to play in the fast-Alfv\'en wave mode conversion process in sunspot atmospheres \citep{cg2008,cally2012}. In a series of theoretical studies using the cold plasma approximation \citep{ca2010,ch2011,hc2011,hc2012}, and more realistic model sunspot atmospheres \citep{kc2012,felipe2012}, it has been demonstrated that conversion to Alfv\'en waves occurs at, and beyond, the fast wave reflection height (located approximately where $c_a \sim \omega/k_h$; where $\omega$ denotes the wave frequency and $k_h$ the horizontal wavenumber), being spread over many scale heights for wavenumbers typical of local helioseismology. The fast-Alfv\'en conversion process appears to be most efficient for $\theta$ (field inclinations from vertical) $\sim$ 30$\degree$ -- 40$\degree$, and $\phi$ (the angle between the magnetic field and wave propagation planes) $\sim$ 60$\degree$ -- 80$\degree$.  

So this leads to a very important question: What are the implications of the returning fast and Alfv\'en waves for the seismology of the photosphere? While the answer to this question currently remains elusive, it does appear that limiting $c_a$, in the manner that we have been doing, could be potentially problematic for active region seismology. 

\section{Conclusion}

It is perhaps not too presumptuous to say that local helioseismology is the greatest benefactor from improved observations provided by SDO. Since 2010, the HMI instrument not only provides higher-resolution Doppler and continuum images of the solar disk, but also high quality line-of-sight and vector magnetograms -- the latter are extremely valuable resources for helioseismologists. Reliable measurements of the photospheric magnetic field are the key to constraining the near-surface layers of MHS sunspot models, as detailed models of the surface layers are a necessity in order to probe the deeper structure of sunspots. Whether the ``monolithic'' (e.g., \citealp{cowling1946,cowling1957,cowling1976}), or ``spaghetti'' (e.g., \citealp{parker1975,parker1979,spruit1981,zwaan1981}) model best explains their subsurface magnetic field structure, still remains to be seen.  

But as we wait for new observational results to roll in, we should, in the meantime, continue to improve the accuracy and efficiency of our numerical forward modelling techniques to address the inconsistencies in the helioseismic inferences of both flux emergence and active region seismology studies. Both areas are also in pressing need of detailed parametric studies, the results from which can be utilised to fine-tune translationally-variant sensitivity kernels  \citep{hanasogeetal2012}, required for any future non-linear inversions of subsurface sunspot structure and dynamics.

\newpage


\end{document}